\documentclass[12pt]{iopart}
\usepackage{iopams}  
\usepackage{graphicx,psfrag}
\usepackage{xcolor}

\newcommand\dfrac[2]{#1 / #2}
\newcommand\text[1]{\mathrm{#1}}
\def\reals{\mathbb{R}}
\newcommand{\jo}[1]{{\color{black} #1}}

\begin{document}

\title[Diffusion and escape from polygonal channels]{Diffusion and
  escape from polygonal channels: extreme values and geometric
  effects}

\author{Jordan Orchard$^{1}$, Lamberto Rondoni$^{2,3}$,
  Carlos Mej\'ia-Monasterio$^{4}$, Federico Frascoli$^{1}$}

\address{$^1$ Department of Mathematics, 
School of Science,
Swinburne University of Technology, H38, PO Box 218, Hawthorn,
Victoria 3122, Australia}
\address{$^2$ Department of Mathematical Sciences, 
    Politecnico di Torino, 
    Corso Duca degli Abruzzi 24, 
    I-10129 Torino, Italy}
\address{$^3$ 
Istituto Nazionale di Fisica Nucleare, Sezione di Torino, Turin, Italy,}
\address{$^4$
School of Agricultural, Food and Biosystems Engineering,
Technical University of Madrid, Av. Complutense s/n, 28040 Madrid, Spain}

\ead{\mailto{jorchard@swin.edu.au},\mailto{lamberto.rondoni@polito.it},\\\mailto{carlos.mejia@upm.es},\mailto{ffrascoli@swin.edu.au}}
\vspace{10pt}

\begin{indented}
\item[]January 2021
\end{indented}

\begin{abstract} 
  Polygonal billiards are an example of pseudo-chaotic dynamics, a
  combination of integrable evolution and sudden jumps due to conical
  singular points that arise from the corners of the polygons. \jo{Such pseudo-chaotic behaviour, often characterised by an algebraic separation of
  nearby trajectories, is believed to be linked to the wild
  dependence that particle transport has on the fine details
  of the billiard table.  Here we address this relation through a detailed numerical study of
  the statistics of displacement in a family of
  polygonal channel billiards with parallel walls.  We show that transport
  is characterised by strong anomalous diffusion, with a mean square
  displacement that scales in time faster than linear, and with a
  probability density of the displacement exhibiting exponential tails
  and ballistic fronts.  In channels of finite length the distribution
  of first-passage times is characterised by fat
  tails, with a mean first-passage time that diverges when the
  aperture angle is rational.} These findings have non trivial
  consequences for a variety of experiments.
\end{abstract}

\vspace{2pc}

\submitto{JSTAT}
\maketitle


\section{Introduction} 
\label{sec:intro}
The transport properties of assemblies of highly confined particles in
nanochannels greatly differ from those of macroscopic fluid systems.
In confined geometries, the interactions between the particles and the
channel walls occur with frequency and magnitude that are comparable
(when not greater) than those associated with particle-particle
interactions, hindering the realisation of Local Thermodynamic
Equilibrium (LTE) on which the standard thermodynamic laws like Fick's
law of diffusion are based \cite{Kreuzer,Spohn}.  When confinement is
particularly pronounced, the conditions for the validity of kinetic
theory fail as well.  For instance, macroscopic containers may keep a
gas in different equilibrium states even if they are connected by
channels that allow the flow of particles but prevent the interactions
among them due to geometry, like in Knudsen gases
\cite{DeGM,CercIllPul}.  Geometry effects due to confinement may also
prevent the decay of microscopic correlations that is required for
locality to hold, allowing hydrodynamic interactions that commonly
lead to an enhancement in the fluid flow
\cite{Thomas:2009bf,Su:2012ey,Jepps:2006be,Jepps:2008ji,Sanders:2006dx}.

When the dimensions of the container are comparable to the molecular
characteristic scales, transport drastically slows down, with
single-file diffusion being a prototypical example
\cite{Levitt,Mon:2002cg,Berezhkovskii:2002fq}.  There are other cases
where instead densely packed systems yield a strong enhancement of the
rate of diffusion \cite{Benichou:2013bm,Illien:2013gx}, and where
anomalies arise as the result of collective mediated interactions due
to confinement, external driving and macromolecular crowding
\cite{MejiaMonasterio:2011kf,Vasilyev:2017im}.  Colloidal systems out
of equilibrium and small biological systems
\cite{Hofling:2013bk,Metzler:2016ju} share similar features.

A prototypical example of transport in confined geometries is that of
diffusion through narrow channels. Understanding whether transport
through nano channels is diffusive or anomalous has gained increasing
interest as it would allow the control of a wide variety of phenomena
and applications, such as ion transport across the cell membrane
\cite{Bruno:2018ev}, drug delivery \cite{Sabek:2013jg}, nanopore
sensing \cite{Reynaud:2020by}, ion transport control for energy
harvesting \cite{Xiao:2019kw}, water flow in carbon nanotubes
\cite{Meersmann:2000cg}, diffusion of waves in corrugated channels
\cite{Aschieri:2006cq}, single photons in optic fibers \cite{SPOF}
among many others.

Denoting by $x(t)$ the position of a tagged particle at time $t$,
transport of matter is customarily defined in terms of the
fluctuations of the displacement $\Delta x= x(t) - x(0)$ of the
particles constituting the fluid.  Macroscopic transport within
stationary media is commonly diffusive, and the displacement is
described by the diffusion equation, whose solution for a localised
initial density is a time dependent Gaussian distribution 
\begin{equation}\label{eq:norm_dis}
    P(\Delta x, t) = \dfrac{1}{\sqrt{2\pi d D t}} \exp \left(-\dfrac{\Delta x^2}{2 d Dt} \right),
\end{equation}
where $d$ is the spatial dimension, and $D$ is the diffusion
coefficient. The size of the fluctuations of the particles'
displacement is then given by the variance of the
displacement, {\em i.e.}  the Mean Square Displacement (MSD). In the
case of Eq.~(\ref{eq:norm_dis}), the MSD is given by:
\begin{equation} \label{def:MSD}
    \langle (x(t) - x(0))^2 \rangle = 2 dD t \ .
\end{equation}
If the MSD is not described by Eq.~\eref{def:MSD}, then transport is
called anomalous. One of the aims of the present study is to
characterise the anomalous transport associated with a specific family
of billiards and discuss the role of geometry in a number of important
properties that describe the billiards' dynamics. 

In this paper we revisit the problem of particle transport in
polygonal billiard channels. Billiards constitute an attractive basis
to study a broad range of fundamental problems in statistical
mechanics, in particular transport in confined geometries
\cite{Sinai:2007kn,Balint:2020vo}. Mathematically, a billiard is a
$d$-dimensional dynamical system, representing one point particle that
moves freely inside a compact domain, the so-called table
$\mathcal{B}\subset\reals^d$, and that is specularly reflected at the
boundary $\partial\mathcal{B}$ of $\mathcal{B}$. The boundary is
required to be piecewise smooth, except in a countable number of
points. Multiple extensions of this simple scenario have been
considered, including interacting particles
\cite{Sinai:2007kn,Lansel:2006dw}, time-dependent boundaries
\cite{Loskutov:2007ki}, thermostated dynamics
\cite{Moran:1987wf,Lloyd95}, stochastic boundaries
\cite{Chumley:2016es}, and boundary with holes \cite{{Demers:2009gh}}.

Normal diffusive transport as described by (2) is usually encountered
in dispersing and semi-dispersing billiards \cite{Binum1,ChernovBuSi},
whose dynamics is characterised by positive Lyapunov exponents. In
constrast, polygonal billiards have zero Lyapunov exponents.  The
separation of nearby trajectories, which is always slower than
exponential, is due to the corners of the polygon associated to the
existence of singular conical points in the billiard flow.  The question
on whether diffusive transport can be obtained in polygonal channels
was \jo{thoroughly} analysed in several works
\cite{Jepps:2006be,Jepps:2008ji,Sanders:2006dx,Alonso:2004bf,Vollmer:2021ut}.
Transport was found to range from diffusive to ballistic, with abrupt
and unpredictable variations of its details under variations of the
geometric parameters.

We focus on those fine details for a particular family of polygonal
billiard channels with parallel boundaries defined in
Section~\ref{sec:poly}. The statistics of the particle displacement is
analysed in Section~\ref{sec:displ}. Later in Section
\ref{sec:jordan}, we analyse the transmission of particles through
channels of finite length and discuss the statistics of survival
probabilities and first-passage times to the channel's exits. Finally
in Section~\ref{sec:concl} we report our conclusions. 
\jo{Our main results are on the interplay between ballistic fronts
and the wild dependence that transport properties have on the geometric parameters. 
Additionally, we show that the distribution of first passage times for channels of finite width is broad, with a mean 
first passage time that diverges for simple rational billiards.}

\section{Polygonal billiard channels}
\label{sec:poly}

\begin{figure}[t]
    \centering
    \includegraphics[width=0.9\textwidth]{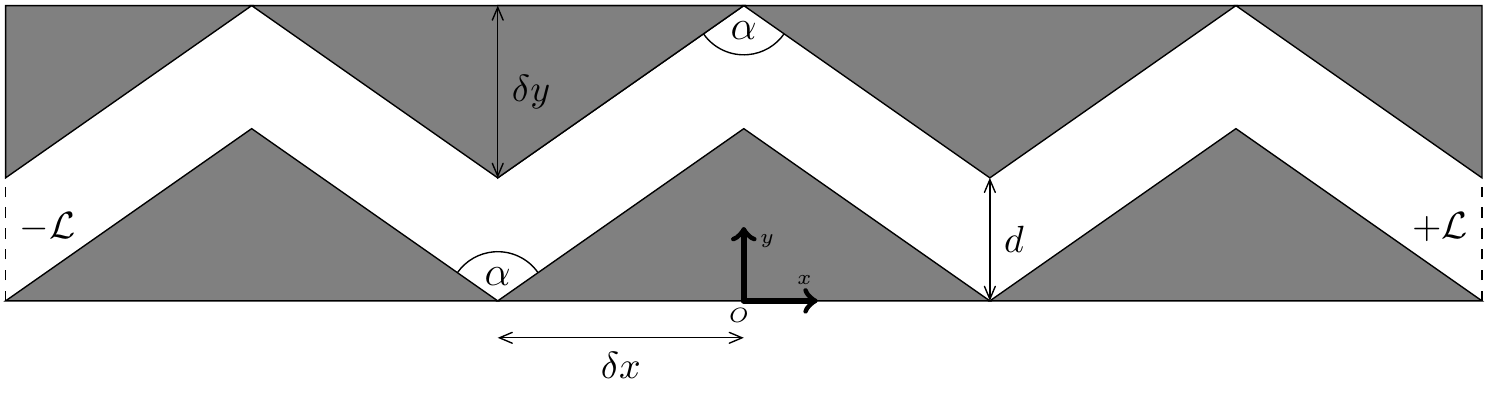}
    \caption{Polygonal channel made of (three) identical copies.  The
      geometry of the fundamental cell is determined by the aperture
      angle $\alpha$ and the cell width $d$, with a cell length fixed
      to $2 \delta x = 1$. The boundary height is given by
      $\delta y = \delta x \cot \dfrac{\alpha}{2}$. Setting the
      central cell at the origin, a channel made of $L$ cells extends
      to $x=\pm \mathcal{L}$, with
      $\mathcal{L} = \lfloor \frac{L}{2} \rfloor + \frac{1}{2}$, and
      $\lfloor\cdot\rfloor$ being the floor function.}
    \label{fig:ChannelSchematic}    
\end{figure}

We consider periodic polygonal billiard channels, made of identical
copies of a fundamental cell with parallel walls, vertically shifted
of a distance $d$ with respect to each other, as shown in
Fig.~\ref{fig:ChannelSchematic}. The fact that the walls are parallel
is an important feature of our systems and will have interesting
consequences, as will be seen shortly.

For all geometries, the horizontal width of the cell is set to
$2\delta x=1$. Therefore, two parameters suffice to specify the shape
of the channel: the vertical shift $d>0$ and the aperture angle
$\alpha\in(0,\pi)$.  When the channel width
$d > \delta x \cot(\alpha/2)$ the horizon is infinite, meaning that a
particle trajectory can move for arbitrarily long distances without
hitting the channel boundaries. Inside the channel, point particles
move with constant speed along straight lines until they elastically
collide with the boundary, where their trajectories are specularly
reflected.

\jo{As noted earlier}, the dynamics of polygonal billiards have zero Lyapunov exponents.
Whether this should imply faster or slower transport is not
obvious. In the case of dispersive billiards, which have positive
Lyapunov exponents, a well-understood example is the Lorentz gas,
which was intensively investigated \jo{a few decades ago}. In that system,
transport can be understood in terms of unstable periodic orbits
\cite{CviGasShr,MorRon}, which subdivide in confined and ballistic
orbits, the latter being periodic only modulo the periodic lattice
spacing.  Confined orbits do not contribute to transport, whereas
ballistic ones do: their combined contribution results in normal
diffusion if the horizon is finite. The strict hyperbolicity of the
system causes generic trajectories to always come close to some
periodic one, but only for a short time in the vicinity of a given
periodic orbit, thus efficiently mixing velocities.

Contrary to the \jo{Lorentz} gas, intuition on polygonal channels' dynamics
is not straightforward, one reason being that parallel walls cause
families of bouncing ball modes to be present. These are periodic
trajectories that collide with the same pair of parallel wall
segments, giving \jo{a} vanishing contribution to transport.  Differently
from the chaotic case, escape from a narrow neighbourhood of a
bouncing ball orbit is very slow and jeopardises the mixing of
velocities. \jo{Only near the polygon's corners trajectories are
separated, and velocities substantially change. 
This implies that trajectories initially in the neighbourhood of a bouncing ball orbit, move slowly away from that orbit, hence, at least initially, they give a small contribution to the transport properties. This behaviour is more prominent in rational polygons, where the number of possible
velocity orientations is finite, {\em e.g.}  just four for
$\alpha=\pi/2$.} A similar phenomenon, however, occurs for the ballistic orbits, {\em
  i.e.}\ those that are periodic only up to periodic lattice vector
translation. These trajectories are very fast, and do not quickly
decorrelate from orbits that start in their vicinity, which overall
results in a large contribution to transport. Which of the two effects
dominates transport and how it is related to the geometry of the
system is a hard open question, which we partly unravel in this paper.

What one observes is that the velocity vector explores a larger set of
discrete orientations in irrational polygons than in rational
ones. With parallel walls, the set of orientations of the particle
velocity does not contract or expand upon successive collisions with
the billiard boundary, as they instead do with non parallel walls.
This leads to a propensity of sets of trajectories to persist along
highly correlated unidirectional flights.  While it is not clear
whether the set of orientations is dense, the exploration of possible
orientations is instead extremely slow
\cite{Gutkin:1996hc,Jepps:2006be}.

This polynomial separation of trajectories  inspired a simple map, the
Slicer  Map  \cite{Salari:2015fx},  to understand  the  properties  of
transport.  The   slicer  map  exhibits  strong   anomalous  diffusion
\cite{Castiglione:1999wz}  and,   as  recently  shown,  the   map  and
polygonal billiard channels belong to a universality class of dynamics
asymptotically      dominated      by      a      ballistic      front
\cite{Vollmer:2021ut}.  It  is  important to  note  that  trajectories
travelling  arbitrarily   long  distances  before  reversing   do  act
effectively  as  ballistic  fronts.  Interestingly,  {such}  ballistic
fronts  {do not  require an}  infinite horizon  but exist  with finite
horizon as in the case  of periodic dispersive billiards.  But, unlike
that case, they  are only weakly unstable because  collisions with the
boundary are  not defocusing: the  result is an overall  poorly mixing
dynamics  that  is  much  harder to  characterise  and  globally  more
unpredictable than chaotic dynamics \cite{Jepps:2006be}.

One of our main goals in this work is to study the statistics of
particle displacement and its relation with the two free parameters
$d$ and $\alpha$ that determine the billiard's geometry.  There are
two main characteristics of the geometry that have a strong impact on
transport: \jo{whether the angle $\alpha$ is a rational or irrational multiple of $\pi$ and
whether the horizon is finite or infinite. Hereafter, when referring to $\alpha$ as being rational or irrational, we will mean that $\alpha / \pi$ is rational or irrational.}  Unless otherwise
specified, for each rational angle
$\alpha_{\mathrm{R}} = \frac{p}{q}\pi$, with $p,q\in\mathbb{Z}$ we
will consider its irrational counterpart defined as
$\alpha_{\mathrm{I}} = \left(\varphi - \frac{3}{5}\right)
\alpha_{\mathrm{R}}$ where $\varphi = \left(\sqrt{5} + 1\right)/2$, so
that $\alpha_{\mathrm{R}}$ and $\alpha_{\mathrm{I}}$ differ by less
than $2\%$.

Furthermore and unless otherwise specified, we consider only two
different channel widths: $d = 2 \delta y$ for the infinite horizon
and $d = \dfrac{\delta y}{2}$ for the finite horizon case, where
$\delta y = \delta x \cot \dfrac{\alpha}{2}$ is the geometric width of
the polygonal boundary (see Fig.~\ref{fig:ChannelSchematic}). When
$\delta y = 1$ the $y$-coordinate of the lower corners located at the
integer values $x=n$ coincide with that of the upper corners at
$x = n+1/2$ and the horizon is critical.

\section{Particle displacement}
\label{sec:displ}

In this Section we study the statistics of the displacement
$\Delta x(t) = x(t) - x(0)$ for noninteracting particles moving in a
parallel polygonal channel of infinite length. At the initial time,
$t=0$, the particles positions $(x(0),y(0))$ are uniformly distributed
inside the central cell, and the velocity vectors {all have}
unit magnitude, with {their orientation forming an angle with the
  horizontal} direction {that is} uniformly distributed over
$[0,2\pi)$.

\subsection{Probability distribution}
\label{sec:pdf}

We focus first on the probability distribution function of the
particle displacement $P(\Delta x,t)dt$, which is the probability that
{one} particle is found at a distance $\Delta x$ from its initial
position and at a time between $t$ and $t+dt$. We have numerically
reconstructed $P(\Delta x,t)$, by {computing} the displacement
$\Delta x(t)$ at different times $t$ {for an ensemble of particles.}
The shape of the probability density $P(\Delta x,t)$, and thus the
properties of the particle transport in the channel, depend on the two
free parameters $d$ and $\alpha$.

Transport {is called scale invariant when} the  probability  density 
scales in time as:
\begin{equation} \label{eq:scaleinv}
P(\Delta x,t)  =  \frac{1}{t^{\gamma}} P\left(\frac{\Delta x}{t^\gamma}\right)
\end{equation}
with $\gamma$ a positive fixed parameter. Here $t^{-\gamma}$ defines
the proper scaling length of the process.  For processes with constant
speed, $\gamma$ {must take} values in $[0,1]$. Then, a generalised
diffusion coefficient can be defined as:
\begin{equation} \label{Eq:genD}
  \mathcal{D} := \lim_{t\to\infty}
  \frac{\mathrm{Var} \left[\Delta x(t)\right]}{2 d t^{2\gamma}} \ ,
  \quad \mathrm{Var} \left[\Delta x(t)\right] = \langle \Delta
  x^{2}(t)\rangle - \langle \Delta x(t)\rangle^2.
\end{equation}  
Normal diffusion corresponds to $\gamma=1/2$, sub-diffusion {to}
$0 < \gamma < 1/2$ and super-diffusion {to} $1/2 < \gamma < 1$. For
$\gamma=1$ the transport is {called} ballistic.
As a consequence of
scale invariance, {the} moments of the displacement obey the
  following relation:
\begin{eqnarray} \label{eq:moments}
\left\langle | \Delta x(t) |^p \right\rangle & = & \int_{-\infty}^\infty | \Delta x |^p
 P(\Delta x,t) \mathrm{d}(\Delta x) \nonumber \\
& = & \int_{-\infty}^\infty | \Delta x |^p P\left(\frac{\Delta
      x}{t^\gamma}\right)
      \mathrm{d}\left(\frac{\Delta x}{t^\gamma}\right) \nonumber \\
& = & t^{p\gamma} \int_0^\infty | u |^p P(u) \mathrm{d}u \ ,
\end{eqnarray}
where $u = \Delta x/t^\gamma$. Assuming that the distribution $P(u)$
is not broad, so that all its moments exist, the integral in the last
line of Eq.~\eref{eq:moments} is finite and we obtain 
\begin{equation} \label{eq:spectrum}
\left\langle | \Delta x(t) |^p \right\rangle \sim t^{p\gamma} \ .
\end{equation}
In particular, the mean-square displacement scales in time as
$\left\langle \left(\Delta x(t)\right)^2 \right\rangle \sim
t^{2\gamma}$.

\begin{figure}[!ht]
    \centering
   \includegraphics[width=0.8\textwidth]{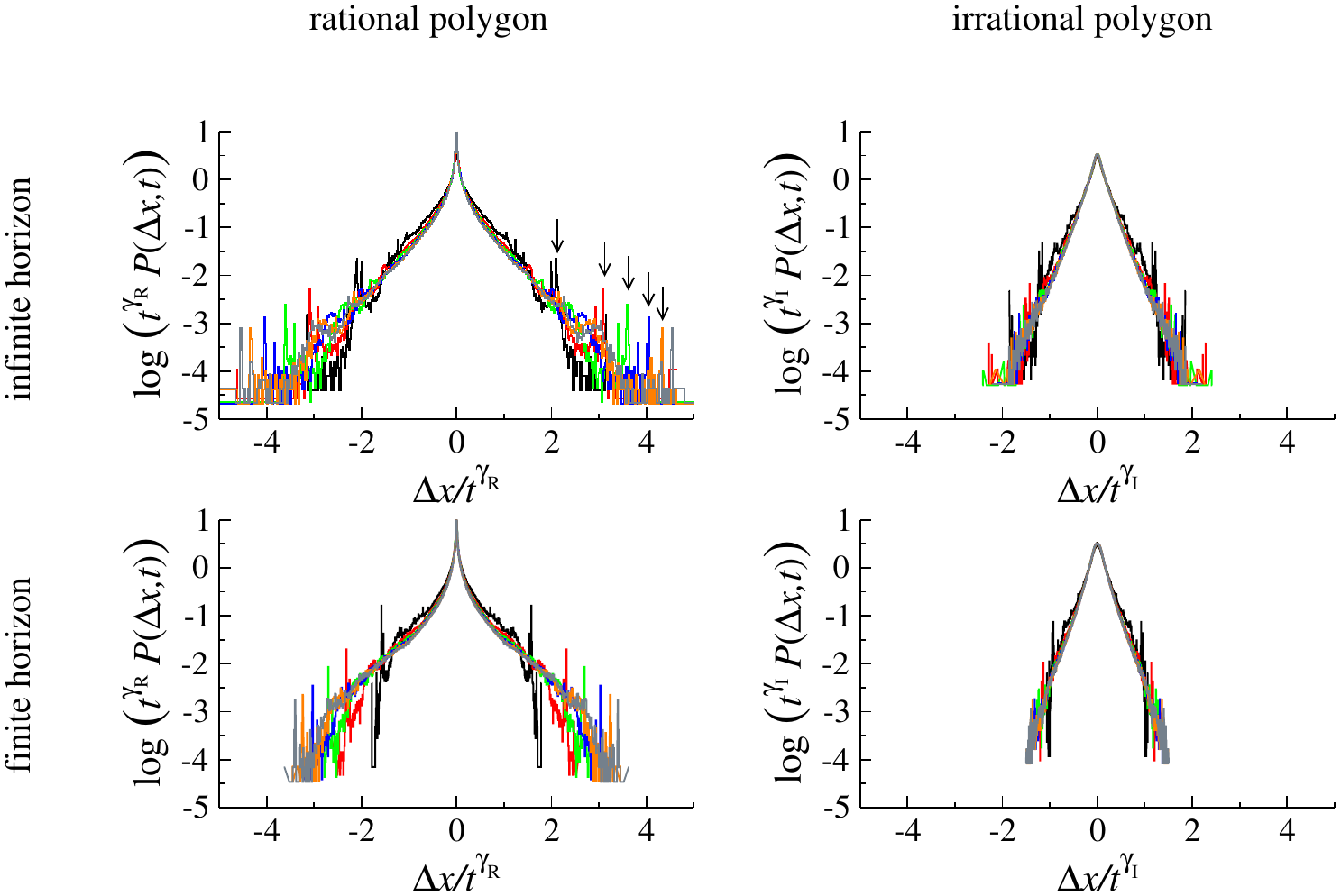}
   \caption{Probability density of the particle displacement $P(\Delta x,t)$
     multiplied by $t^\gamma$, plotted as a function of scaled
     variable $\Delta x/t^\gamma$ obtained from $10^6$ particle
     trajectories. Each histogram corresponds to $t=1000$ (black),
     $10000$ (red), $25000$ (green), $50000$ (blue), $75000$ (orange),
     and $100000$ (grey). The probability density is shown for four
     different billiards as indicated by the labels: \jo{rational polygon $\alpha_R = {\pi}/{2}$ and the corresponding irrational polygon $\alpha_I$}, 
     finite horizon with $d = {\delta y}/{2}$ and infinite horizon for
     $d = 2 \delta y$. The scaling used was $\gamma_R=5/6$ for the
     rational polygon and $\gamma_I=0.91$ for the irrational polygon.
     The arrows in the upper left panel indicate the position of the
     ballistic front at different times (see
     Fig.~\ref{fig:dis_ballistic}).}
    \label{fig:dis_density}
\end{figure}

In order to analyse the scaling of the probability distribution
function of the particles' displacement, we have numerically followed
an ensemble of $10^6$ trajectories, for a time $t=10^5$, for four
different polygonal channels: a rational polygon with
$\alpha_R={\pi}/{2}$ with finite $(d = \delta y/2)$ and infinite
$(d = 2 \delta y)$ horizons, as well as an irrational polygon with
$\alpha_I = \frac{\pi}{2}\left(\varphi-\frac{3}{5}\right)$, for the same
wall separation $d$. In Fig.~\ref{fig:dis_density} we show  the probability distribution of
the   particle  displacement   computed   at   six  different   times:
$t  =  \{1000,10000,25000,50000,75000,100000\}$.   Within  the  fairly
large  space and  time scales  that we  have explored,  we found  that
$P(\Delta  x,t)$ is  indeed not  broad, and  has exponential  or multi
exponential tails,  that largely dominate the  distribution. Moreover,
$P(\Delta  x,t)$  is  also  scale-invariant  and,  interestingly,  the
scaling  {parameter}  $\gamma$  substantially   depends  only  on  the
aperture angle $\alpha$,  whereas its dependence on $d$  appears to be
negligible.  \jo{We  find that the scaled  distributions do asymptotically
collapse      into     the      same     master      function     with
$\gamma=\gamma_R\approx  {5}/{6}$  and $\gamma=\gamma_I\approx 0.91$ for the rational and irrational polygons defined above. In the following section, we will consider the dependence on the aperture angle in more detail.}

At  finite  times  and  large   displacements,  we  observe  that  the
distribution contains  structures that are characteristic  of ballistic
fronts, with peaks of higher probability  that move away \jo{from} the center
as time  elapses.  Interestingly,  these ballistic fronts  are present
for  all  four  cases,  regardless  of the  horizon  being  finite  or
infinite. The position of the  ballistic fronts is indicated by arrows
in the upper left panel of Fig.~\ref{fig:dis_density}.

The  existence  of   ballistic  fronts  means  that   the  scaling  of
Eq.~\eref{eq:scaleinv} has to be complemented by a second contribution
$\mathcal{F}(\Delta x,t)$ such that
\begin{equation} \label{eq:ball}
  \mathcal{F}(\Delta x,t) = \frac{1}{t} \mathcal{F}\left(\frac{\Delta
      x}{t}\right) 
\end{equation}
satisfies      a      ballistic     scaling      (see      \emph{e.g.}
Ref.~\cite{Burioni:2010jp}). Clearly,  $\mathcal{F}(\Delta x,t)$  is a
sub leading   contribution   describing    the   behaviour   at   large
displacements,  \emph{i.e.}    \jo{$t^\gamma \ll  \Delta   x  <   t$},  with
appropriate scaling velocities.  The collapse described by the leading
scale-invariant   contribution  of   Eq.~\eref{eq:scaleinv}  is   then
observed for displacements $\Delta x/t^\gamma$ smaller than the scaled
position of the ballistic front, as expected.

Furthermore,  we also  observe that  the {bulk}  of the  distributions
{are} different:  while for the  rational polygons the centers  of the
distributions {are} spikelike, for  the irrational polygons {they are}
rounded. We notice that the bulk  of the distribution, which in scaled
units is defined  by $\Delta x/t^\gamma < 1$, has  the same dependence
as its tails.  These  quantitative and qualitative differences between
rational and  irrational polygons have  been found for  all geometries
considered here.

Looking at the evolution of the distributions one intuitively realises
that in polygons with infinite horizon the distribution spreads more
rapidly than those with finite horizon, because the progression in
time is essentially determined by the speed of the ballistic front.

\begin{figure}[!ht]
    \centering
    \includegraphics[width=0.8\textwidth]{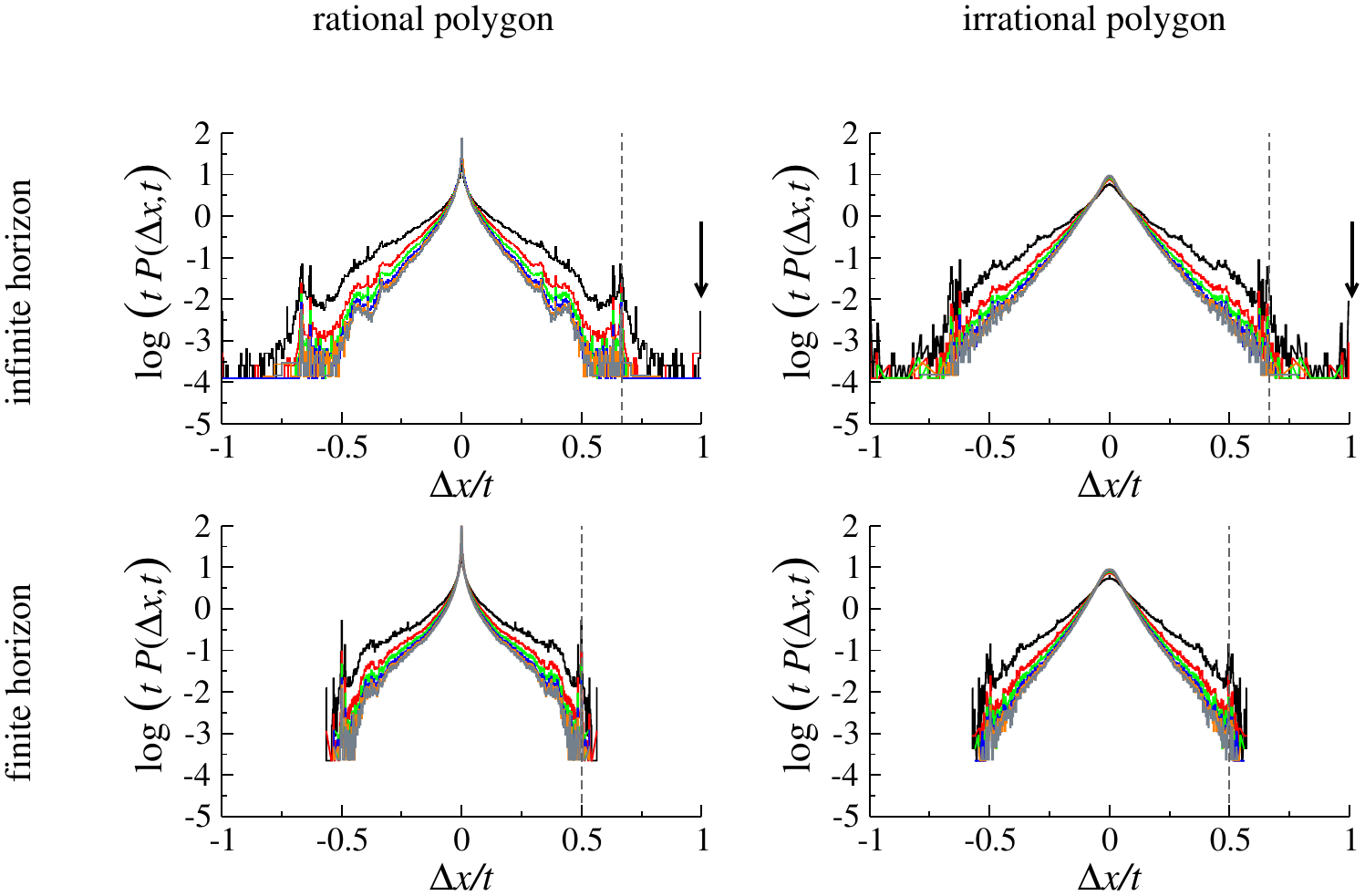}
    \caption{Ballistic scaling of the probability density of the
      particle displacement $P(\Delta x,t)$ shown in
      Fig.~\ref{fig:dis_density}.  The dashed vertical lines are a
      guide to the eye, indicating the value of the speed of the
      ballistic front $v_{bal}=\Delta x/t$. The arrow in the upper left panel
      indicates the non-collisional ballistic front with unit speed.}
    \label{fig:dis_ballistic}
\end{figure}

In Fig.~\ref{fig:dis_ballistic} we plot the \jo{probability
density} of the particle displacement using a ballistic scaling,
obtained with an exponent $\gamma=1$. The scaling variable $\Delta x/t$
corresponds in this case to the speed of the front that we denote as
$v_{\mathrm{bal}}$: this is useful to distinguish the features of the
distribution that scale ballistically.

Let us discuss first the rational polygon (left column of
Fig.~\ref{fig:dis_ballistic}). In scaled units $\Delta x/t$ the position of
the different peaks indicated {by the} arrows in
Fig.~\ref{fig:dis_density} do collapse (as indicated in
Fig.~\ref{fig:dis_ballistic} by the vertical dashed lines), thus
confirming that {they} correspond to a ballistic front.  For the
rational polygon with infinite horizon, the speed of the ballistic
front is approximately $v_{\mathrm{bal}}\approx2/3$.  With finite
horizon, the existence of a ballistic front is also confirmed, with a
different speed $v_{\mathrm{bal}}\approx1/2$.  Moreover, the fastest
widening trajectories with finite horizon correspond precisely to the
ballistic front, {while in the case of} infinite horizon there exist
trajectories that spread faster: those that undergo no collisions and
have speed $\Delta x/t = 1$ (indicated by the arrow in the upper left panel
of Fig.~\ref{fig:dis_ballistic}).  Without acceleration, no trajectory
can be faster than this free-flight front, which arise as soon as the
channel's horizon is infinite and independently of the angle $\alpha$.
In addition, the fraction of such fastest trajectories decreases with
$t$, since it approaches the set of exactly horizontal trajectories.

Turning our attention to the irrational polygon, we do indeed see the
free-flight ballistic front when the horizon is infinite, in addition
to the slower ballistic front. The latter varies with the
geometry of the billiard, but it exists whether the horizon is finite
or infinite. We speculate that this ballistic front is due to the
trajectories that initially are in a close neighbourhood of the
shortest periodic orbit of the billiard. As such, other fronts are
expected to exist and indeed we observe certain features of the
probability distribution that collapse in position under the ballistic
scaling. This may also explain the different velocity of the front for
different geometries of the billiard. Given the differences between
rational and irrational polygons, it comes as a surprise that the
front speed seems to depend strongly on the channel's width, but
apparently {only} mildly on whether the polygon is rational or not.

\begin{figure}[!t]
    \centering
   \includegraphics[width=0.8\textwidth]{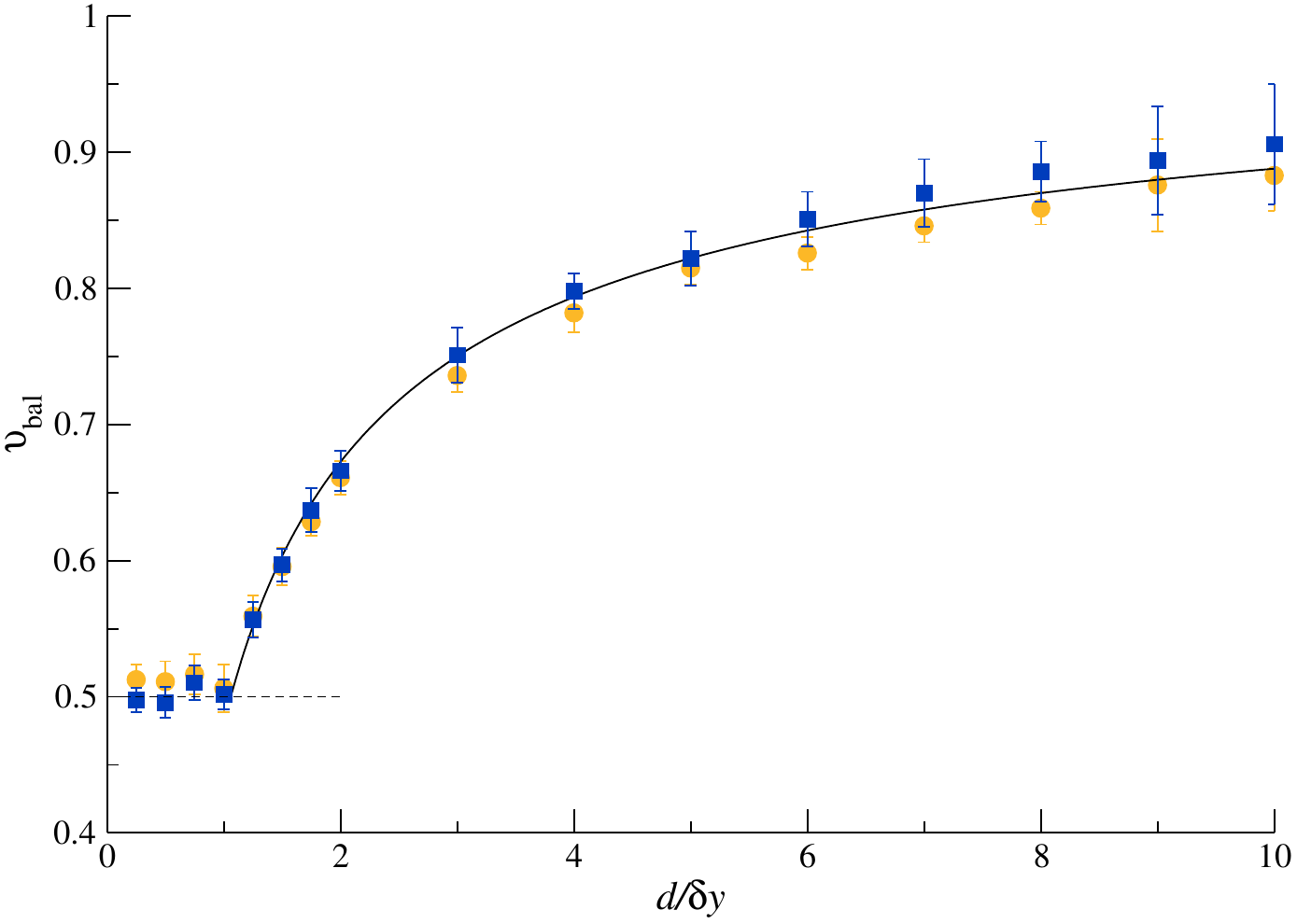}
   \caption{Velocity of the ballistic front as a function of the width
     of the channel $d$ in units of $\delta y$ for the irrational
     angle $\alpha = \frac{\pi}{2}\left(\phi-\frac{3}{5}\right)$
     (light yellow circles) and the rational angle
     $\alpha = \frac{\pi}{2}$ (dark blue squares). The dashed line
     indicates the value $v_{\mathrm{bal}} = 1/2$. The solid curve corresponds
     to the function $v_{\mathrm{bal}} = 1 - 0.52 (d/\delta y)^{-2/3}$.}
    \label{fig:dis_speed}
\end{figure}

To study this further, we have numerically evaluated the speed of the
ballistic front for the rational $\alpha=\pi/2$ polygonal channel and
its irrational counterpart, for different widths. The results are
shown in Fig.~\ref{fig:dis_speed}, where the dark blue squares refer
to the rational polygon, and the light yellow circles to the
irrational one.  Opening the horizon leads to a sharp transition in
the behaviour of the speed of the ballistic front: for finite horizon
$d<\delta y$, the speed of the ballistic front does saturate to
approximately $1/2$. When the horizon opens, the speed
$v_{\mathrm{bal}}$ grows monotonically toward its upper bound
$v_{\mathrm{bal}}=1$. From a fit to power-law, we obtain that the
speed approaches unity as
\begin{equation} \label{eq:speed} 
v_{\mathrm{bal}} = 1 - 0.52 \left(\frac{d}{\delta y}\right)^{-2/3} \ , \ \ \mathrm{for} \quad
\frac{d}{\delta y} > 1 \ ,
\end{equation}
as shown by the solid curve in Fig.~\ref{fig:dis_speed}.  As a final
remark we note that the ballistic front of the polygonal channel is
mildly affected by whether the angle is rational or
irrational. However, recalling that
$\left(\varphi - \dfrac{3}{5}\right)\approx 1$, and that the angles
compared in Fig.~\ref{fig:dis_speed} are very similar in value, we
{cannot} discard {the possibility} that $v_{\mathrm{bal}}$ could
depend on the value of the aperture angle in a stronger
fashion. Moreover, conjecturing that the speed of the ballistic front
depends on the length of the simplest periodic orbit,
$v_{\mathrm{bal}}$ should depend evenly on all the parameters of the
geometry of the billiard.

\subsection{Mean-square displacement}
\label{sec:MSD}

In the previous section we have analysed the properties of the
probability distribution of the particle displacement and established
that transport is anomalous, with $P(\Delta x,t)$ having exponential
tails and a super-diffusive scaling, with an exponent that depends on
whether the angle is rational or not.  In this section we further
explore the dependence of transport on the aperture angle $\alpha$,
focussing on the mean-square displacement. For a scale-invariant
probability distribution, the mean-square displacement scales
asymptotically as a power-law of time
$\langle \left(\Delta x(t)\right)^2\rangle \sim t^{\eta}$, where the
exponent $\eta$ is related to the scaling exponent of the probability
distribution as $\eta=2\gamma$ (see Eq.~\eref{eq:spectrum}).

\begin{figure}[!ht]
    \centering
    \includegraphics[width=0.8\textwidth]{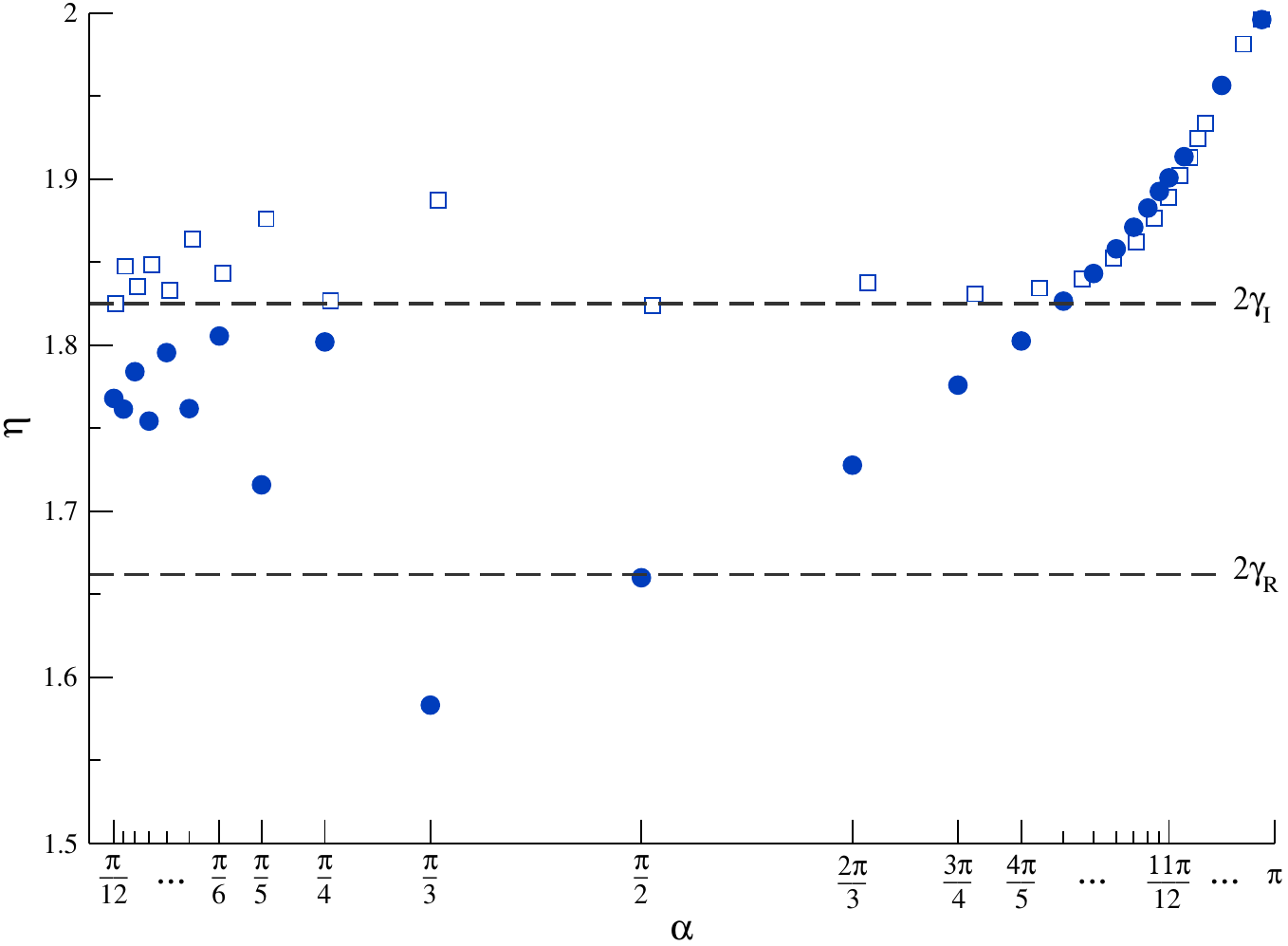}
    \caption{MSD scalings $\eta$ for rational (solid circles) and
      irrational billiards (open squares). Simulations involved $10^6$
      trajectories of length $t=10^6$ for irrational $\alpha$'s and
      $t=10^5$ for the rational $\alpha$'s.  A finite horizon
      channel with $d = \dfrac{\delta y}{2}$ was used for all cases,
      but a similar behaviour occurs for infinite horizons. The
      horizontal dashed lines indicate the expected values for the MSD
      for $\alpha=\pi/2$ from the corresponding scalings
      $\gamma_{\mathrm{R}}$ and $\gamma_{\mathrm{I}}$ of the
      probability distribution function. Statistical errors are
      of the same magnitude as the symbols' size.}
    \label{fig:struc_stab}
\end{figure}

We have numerically computed the mean-square particle displacement for
polygonal channels with a finite horizon, $d=\delta y/2$, and
different rational and irrational values of the angle $\alpha$. The
scaling exponent $\eta$ is obtained from a fit to power-law of
$\langle x^2(t)\rangle$ in the large time limit.  The angles were
chosen as rational values between $\pi/12$ {and} $11\pi/12$.

We show in Fig.~\ref{fig:struc_stab} the power exponent $\eta$ of the
mean-square particle displacement for different angles $\alpha$, in
blue crosses rational polygons and in red circles their irrational
counterparts.  For $\alpha=\pi/2$ and
$\alpha = \frac{\pi}{2}\left(\varphi-\frac{3}{5}\right)$ we obtain
$\eta=1.662\pm 0.006$ and $\eta = 1.825\pm 0.006$ respectively, in
agreement with the values of $\gamma_R$ and $\gamma_I$ obtained in the
previous section.

Moreover, for all the parameters considered, the transport in parallel
walls polygonal channels is found to be super-diffusive.  Restricted to
rational polygons, $\eta$ seems to monotonically increase for
$\alpha\ge \pi/3$. At smaller angles the power scaling $\eta$
concentrates at values around $1.78$. We recall that for $\alpha=\pi$
the corners of the polygonal channel disappear and the channel becomes
rectilinear, yielding $\eta=2$. Therefore, we expect that
$\lim_{\alpha\rightarrow\pi} \eta = 2$ irrespective if the angle is
rational or not, as confirmed by Fig.~\ref{fig:struc_stab}. In the
irrational polygons, transport appears to be more stable than in the
rational polygons, with a power scaling concentrated inside a small
interval around $\eta \approx 1.84$.

As a final remark we notice  that the numerical convergence of the MSD
statistics  of  the  irrational  polygons for  small  aperture  angles
$\alpha  \lesssim  \pi/4$,  is  drastically  slower  than  for  larger
angles. The properties of the  convergence to the asymptotic transport
and  how   they  are  affected   by  the  geometry   deserves  further
investigation.

\subsection{Higher order moments}
\label{sec:SAD}

To complete the picture we have numerically computed higher order
moments of the particle displacement
$\langle\left|\Delta x(t)\right|^p\rangle$.

In Section~\ref{sec:pdf} we found that the probability distribution
of the particle displacement is scale invariant (see Eq.~\eref
{eq:scaleinv}), with a scaling $x/t^\gamma$ that depends on the
geometry of the billiard.  The scale-invariance of $P(\Delta x,t)$
exhibiting exponential tails does hold for displacements that are
shorter than the position of the ballistic front, namely in scaled
units for
\begin{equation} \label{eq:bound}
\frac{\Delta x}{t^\gamma} < v_{\mathrm{bal}}  \ t^{1-\gamma} \ .
\end{equation}
Different scales determining the evolution of $P(\Delta x,t)$ commonly
lead to strong anomalous diffusion characterised by a non-homogeneous
scaling of the moments of the displacement with time. Strong anomalous
diffusion, first described in Ref.~\cite{Castiglione:1999wz}, 
is characterised by the relation:
\begin{equation} \label{def:SAD}
\langle\left|\Delta x(t)\right|^p\rangle \sim t^{\nu(p)} \ ,
\end{equation} 
where the scaling exponent $\nu(p)$ is {not linear in} the
moment's order $p$.

Strong anomalous diffusion was observed in Ref.~\cite{Sanders:2006dx}
for general polygonal channels with finite and infinite horizon.
There, the authors found that the moments of displacement can be
subdivided in two branches: a scaling
$\langle\left|\Delta x(t)\right|^p\rangle \sim t^{\nu_{\mathrm{low}}}$
for low order moments $p<p^\star$, and
$\langle\left|\Delta x(t)\right|^p\rangle \sim
t^{\nu_{\mathrm{high}}}$ for $p>p^\star$. This piecewise behaviour is
characteristic of systems with two different dynamic scales
\cite{Castiglione:1999wz} and was recently shown \cite{Vollmer:2021ut}
that polygonal channels with infinite horizon belong to a broader
class of universality of transport that is asymptotically dominated by
ballistic fronts.

\begin{figure}[!t]
    \centering
    \includegraphics[width=0.9\textwidth]{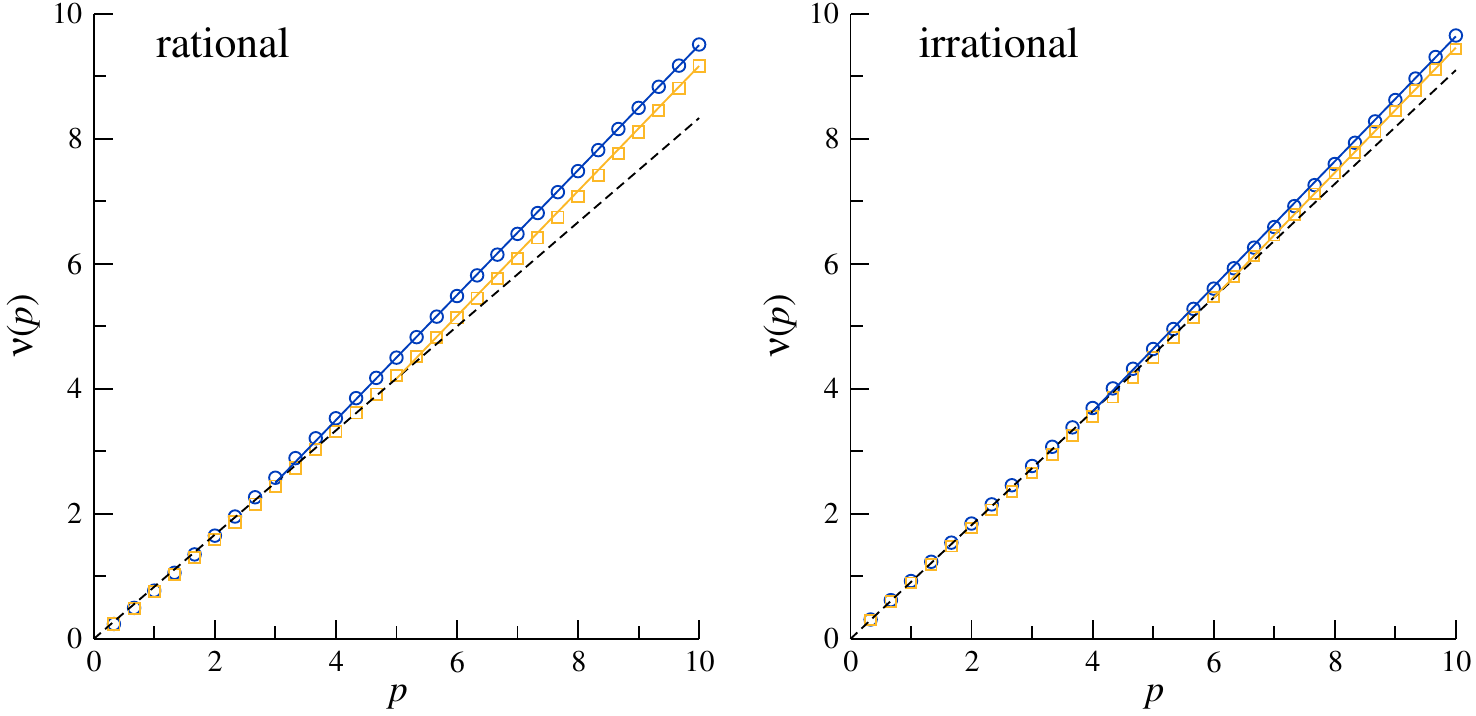}
    \caption{Spectrum of the moments of displacement of
      Eq.~\eref{eq:spectrum} for the same channels as in
      Fig.~\ref{fig:dis_density}, rational polygon with
      $\alpha = {\pi}/{2}$, irrational polygon with
      $\alpha = \frac{\pi}{2}\left(\phi-\frac{3}{5}\right)$, and
      finite horizon with $d = \frac{\delta y}{2}$ (dark blue
      circles), and infinite horizon for $d = 2 \delta y$ (light
      yellow squares). In the left panel the lines are $5p/6$
      (dashed), and $p-1/2$, $p-5/6$ (in solid).  In the right panel
      the lines are $0.91p$ (dashed), and $p-0.36$, $p-0.54$ (in
      solid). Solid lines are shown with the same colour as the
      corresponding symbols.  }
    \label{fig:spectrum}
\end{figure}
 
Here we find that, independently of {the} geometry of the polygon,
transport is strongly anomalous, with a piecewise linear spectrum
characterised by superdiffusion for moments of order $p\le p^\star$
and ballistic branch for higher moments. This is shown in
Fig.~\ref{fig:spectrum}, where we plot the scaling exponent
\begin{equation} \label{eq:spectrum}
    \jo{\nu(p)  = \lim_{t\rightarrow\infty} \frac{\log \langle\left|\Delta x(t)\right|^p\rangle}{\log t}} \ 
\end{equation}
for the same polygonal channels investigated in section~\ref{sec:pdf}.
Consistently with the previous results, low order moments of the
displacement with $p<p^\star$ and $p^\star>2$, scale as
\begin{equation} \label{eq:SAD-1}
\nu(p) = \left\{
\begin{array}{ll}
\frac{5}{6} \ p & \textrm{for the rational polygon}\\
\\
0.91 \ p & \textrm{for the irrational polygon}
\end{array}
\right. \ 
\end{equation}
for arbitrary horizons. Higher order moments scale ballistically, with
a threshold order $p^\star$ that depends on the geometry, as shown in
Table~\ref{tab:spectrum}.  We conjecture that this behaviour is
qualitatively observed for arbitrary values of $\alpha$ and of
$d$. Moreover, Fig.~\ref{fig:struc_stab} shows that for
$\alpha<\pi$ the MSD scales slower than ballistic, which means that
the threshold moment order is $p^\star > 2$ independently of the
geometry.

\begin{table}[ht] 
\centering
\begin{tabular}{|c|c|c|c||l|c|}
    \hline 
polygon & horizon &   $\alpha$    & $d$ & $p^\star$ & $\nu(p)$ \\
    \hline 
    \hline 
rational & finite &     $\frac{\pi}{2}$ & $\delta y/2$ & $3$ & $p  - \frac{1}{2}$ \\
rational & infinite &     $\frac{\pi}{2}$ & $2\delta y$ & $5$ & $p  - \frac{5}{6}$ \\
irrational & finite &    $\left(\varphi-\frac{3}{5}\right)\frac{\pi}{2}$ & $\delta y/2$ & $4$ & $p  - 0.36$ \\
irrational & infinite &    $\left(\varphi-\frac{3}{5}\right)\frac{\pi}{2}$ & $2\delta y$ & $6$ & $p  - 0.54$ \\
    \hline 
\end{tabular}
\caption{Asymptotic ballistic spectrum of the moments of the
  displacement $\nu(p)$ for $p>p^\star$, for the four combinations
  investigated.}
\label{tab:spectrum}
\end{table}

\section{Transmission through polygonal channels of  finite length}
\label{sec:jordan}

In the previous section we have shown that the transport across
parallel walls polygonal channels is super-diffusive and dominated
asymptotically by a (possibly more than one) ballistic front, at
different speeds. In this section we investigate how transport is
determined in channels of finite length for which the asymptotic
regime is not available.

We have explored channels of finite length with ends located at
positions
$x = \pm \mathcal{L} = \left(\lfloor \frac{L}{2} \rfloor + \delta x
\right)$, and act as absorbing boundaries.  We recall that $L \ge 1$ is
the number of polygonal cells composing the channel,
$\lfloor\cdot\rfloor$ refers to the greatest integer that is less than
or equal to the argument, and $\delta x = 1/2$.  When a particle
reaches one of the two ends, it escapes the channel and is removed
from the dynamics.

As particles have constant speed, finite size effects will appear
effectively at times larger than $\tau^\star$, the shortest time at
which particles escape from the channel. If the horizon is infinite
then $\tau^\star=\lfloor\frac{L}{2}\rfloor$ exactly. At finite
horizon, the shortest absorption time is determined by the fastest
ballistic front for the given geometry, and is approximatively
$\tau^\star = \mathcal{L} / v_{\mathrm{bal}}$.

For finite channels, the moments of the particle displacement can be
defined at time $t$ as a conditional average over the particles that
are still inside the channel. For a total number $M$ of particles we
define
 \begin{equation}\label{eq:fmom}
   \langle |\Delta x(t)|^p \rangle = \frac{1}{\mathcal{N}_t}
   \sum_{j=1}^M \chi_j(t) \left|\Delta x_j(t) \right|^p \ ,
\end{equation}
where the indicator function $\chi_j(t)=1$ if by time $t$ the $j$-th
particle has not yet been absorbed, and {$\chi_j(t)=0$
  otherwise. Then}
\begin{equation}
    \mathcal{N}_t = \sum_{j=1}^M \chi_j(t) \ 
\end{equation}
is the total number of particles that by time $t$ have not escaped
from the channel.

\begin{figure}[!t]
    \centering
   \includegraphics[width=0.8\textwidth]{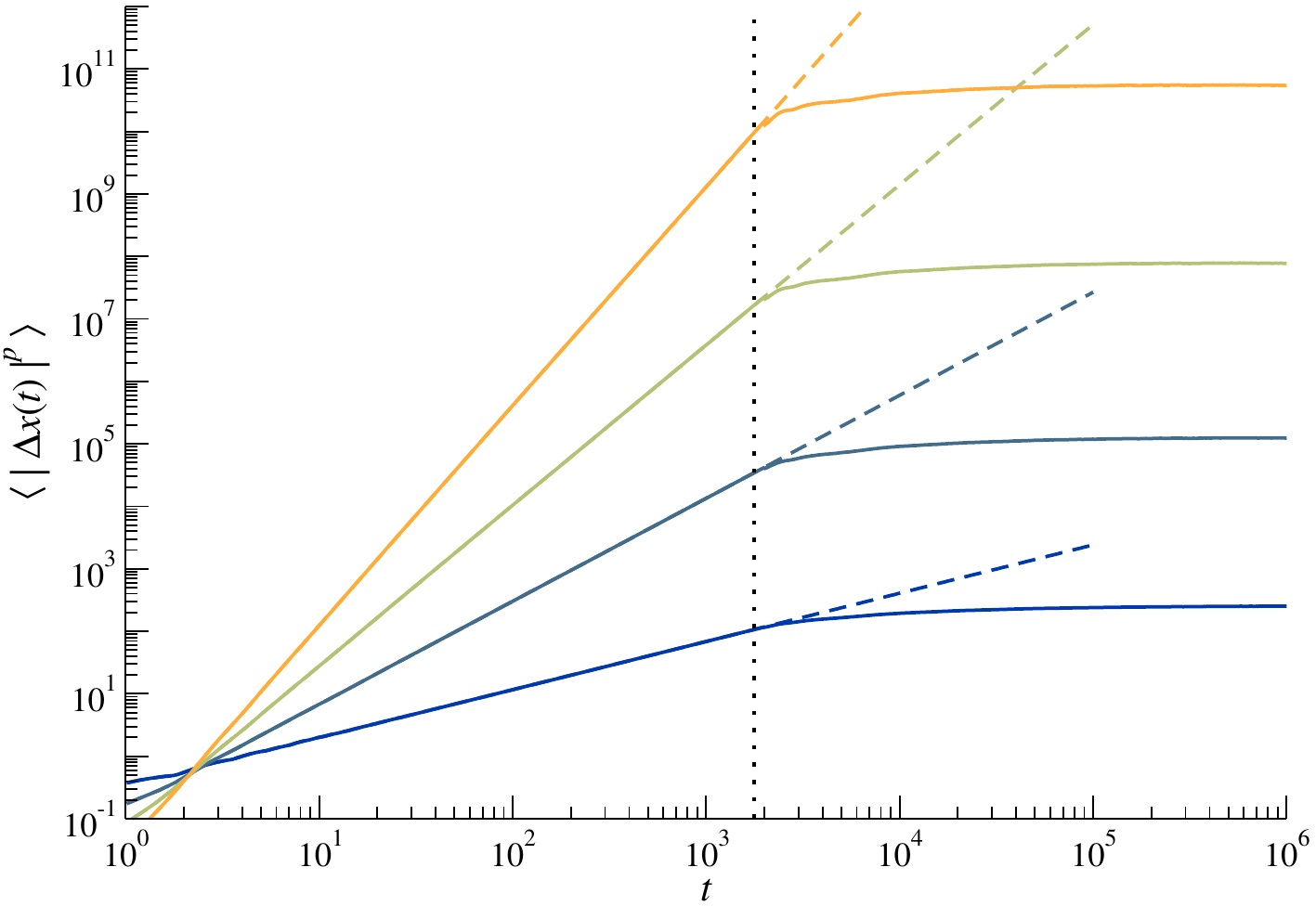}
   \caption{First four moments of the particle displacement
     $p=1,\ldots, 4$ (from dark blue to light yellow), for a channel
     composed of $L = 2001$ cells, $\alpha = \dfrac{\pi}{2}$ and
     $d = \dfrac{\delta y}{2}$ as per Eq.~\eref{eq:fmom} (solid
     curves). The dashed curves correspond to the moments for the
     channel with infinite length. The dotted vertical line indicates
     the value of the shortest first-passage time $t = \tau^\star$.}
    \label{fig:fmom1}
\end{figure}

We have numerically computed Eq.~(\ref{eq:fmom}) for a channel
composed of $L=2001$ rational cells with $\alpha = \dfrac{\pi}{2}$ and
finite horizon $d = \dfrac{\delta y}{2}$. The first four moments are
shown in Fig.~\ref{fig:fmom1}, indicating a growth similar to that of
the infinite channel (shown in dotted lines) upon saturation at
$t=\tau^\star$.

The saturation arises from the definition of the conditional sum in
Eq.~\eref{eq:fmom}, {and} gives quantitative bounds for the moments,
in particular for the mean-square displacement.  In a channel of semi
length $\mathcal{L}$, each particle contributes at most
$|\mathcal{L} - x(0)|^2$ to the MSD: since the remaining particles in
the channel give a smaller contribution, the MSD saturation value is
{of order $O(\mathcal{L}^2)$.}

\subsection{Statistics of first-passage times}
\label{sec:fptd}

The time $\tau^\star$ corresponds to the shortest time at which a
particle escapes the channel.  In this section we explore the
properties of this escape process in terms of the statistics of the
first-passage time.

The concept of first-passage underlies a number of diverse stochastic
processes, {when the} relevant {information is represented by the
  first time at which} the value of a random variable reaches a preset
value \cite{Redner:2007ua}. The escape time of a given trajectory
$x(t)$ can be seen as a first-passage time defined by:
\begin{equation} \label{eq:fpt}
\tau = \inf\{t : |x(t)| = \mathcal{L} \}\ ,
\end{equation}
where the $\inf$ is defined over the set of randomly initialised
trajectories.

Consider a large set of trajectories parametrised by their initial
position $\{(x_j(0),y_j(0))\}$ and initial velocity vector
$\{\vec{v}_j \mathrel{}:\mathrel{}  |\vec{v}_j|=1\}$, and denote by $\tau_j$ the time $t$ at which the
$j$-th particle {exits from one of the two} channel's ends. Then
borrowing the concept from stochastic dynamics, the normalised
distribution $\Psi(\tau)$ of the escape times $\{\tau_j\}$ can be
called the First-Passage Time Distribution (FPTD). 
For a randomly chosen initial condition, $\Psi(\tau)\mathrm{d}\tau$
is the odds that such a trajectory will escape at a time between
$\tau$ and $\tau +\mathrm{d}\tau$.  The mean escape time is then the
Mean First-Passage Time (MFPT) defined as
\begin{equation} \label{def:mfpt}
\langle\tau\rangle = \int_0^\infty \tau\Psi(\tau)\mathrm{d}\tau \ .
\end{equation}

An interesting question is whether or not the Mean First-Passage Time
(MFPT) characterises the typical escape time.  As we learned in the
previous section, the spreading of the trajectories {occurs in}
different {fashions, of which only part are} ballistic.  Therefore, it
is not a priori clear if the MFPT is enough to describe the escaping
process.

\begin{figure}[!t]
    \centering
    \includegraphics[width=0.8\textwidth]{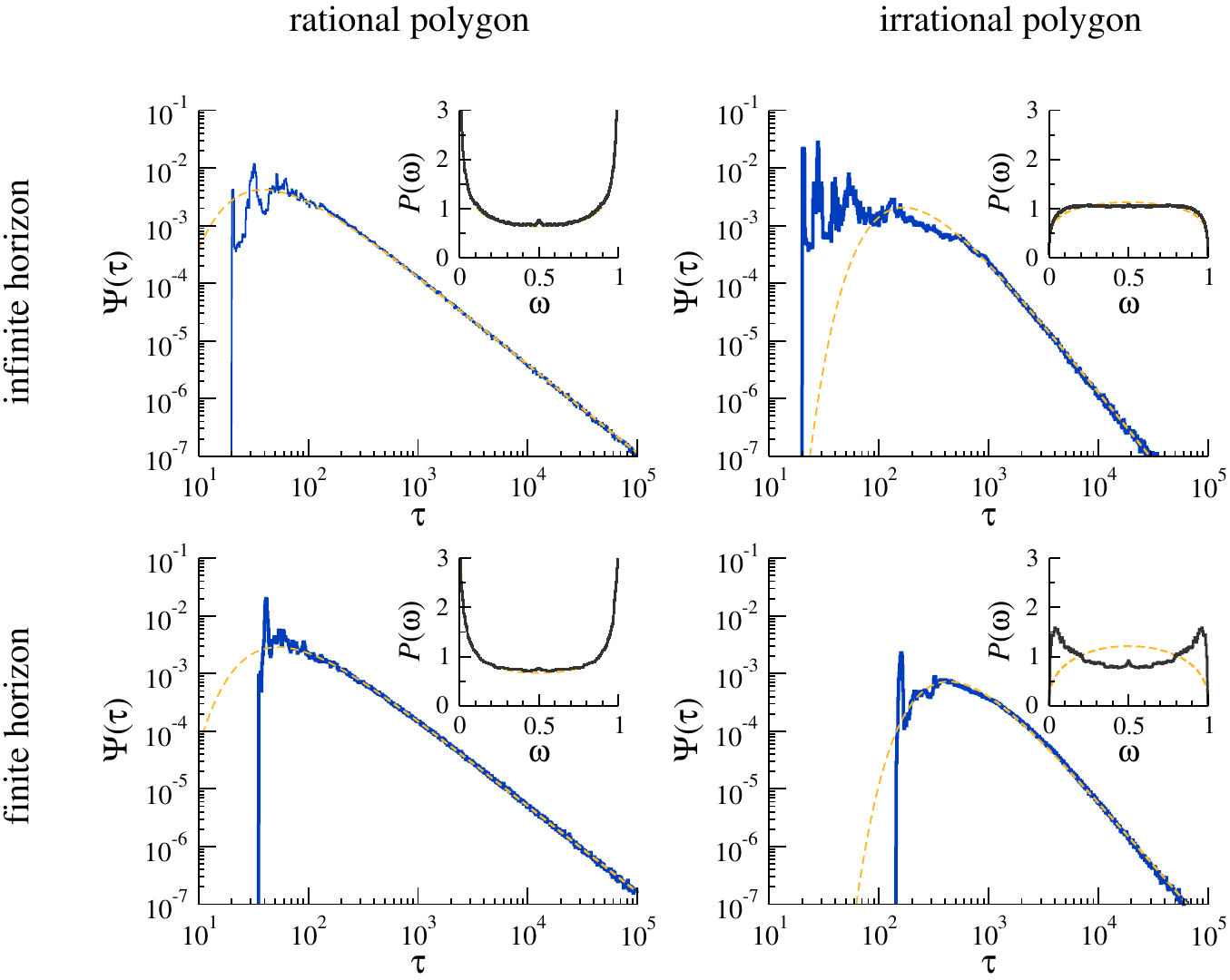}
    \caption{Distribution  of  the   first-passage  time  $\Psi(\tau)$
      (solid dark blue  curves) to exit a finite channel  of $L = 41$,
      for  the  same   geometries  as  in  Fig.~\ref{fig:dis_density}:
      rational  polygon  with  $\alpha  =  \frac{\pi}{2}$,  irrational
      polygon                                                     with
      $\alpha  =  \frac{\pi}{2}\left(\phi-\frac{3}{5}\right)$,  finite
      horizon with $d  = \frac{\delta y}{2}$ and  infinite horizon for
      $d  = 2  \delta  y$.  Light yellow  dashed  curves  are fits  to
      Eq.~\eref{eq:BM}      with       fitting      parameters      as
      in Table~\ref{tab:fitparam}.    Insets:   Numerically   obtained
      distribution  of  the  similarity   index  $P(\omega)$  for  the
      corresponding geometries (solid black  curves). The dashed light
      yellow      curves      correspond     to      Eq.~\eref{eq:Pw}.
    }  \label{fig:fptd}
\end{figure}

We have numerically computed the distribution of the first-passage
time in channels of finite length and the same geometry as those
studied in Section~\ref{sec:pdf}.  As for the infinite channel, a
large set of trajectories are randomly set in the central polygon
cell, and for each of them we measure the time $\tau$ at which each
trajectory reaches the channel ends. The histogram of this set of
times is then used to numerically approximate the FPTD.  The results
are shown in Fig.~\ref{fig:fptd} as solid dark blue curves.

At short times, the distribution $\Psi(\tau)$ has a cutoff at
$\tau=\tau^\star$, namely $\tau^\star=\lfloor\frac{L}{2}\rfloor$
corresponding to the shortest free-flight trajectory in channels with
infinite horizon, while for channels with finite horizon
$\tau^\star=\mathcal{L}/v_{\mathrm{bal}}$, where $v_{\mathrm{bal}}$ is
the speed of the fastest ballistic front.  At later times, we observe
an oscillating behaviour due to the escape of other slower ballistic
fronts.  More importantly, we note that $\Psi(\tau)$ decays
asymptotically as a power-law of $\tau$, meaning that the FPTD is
broad and thus only a finite number of its lowest moments 
exist.

It is physically reasonable to describe our numerical results for
$\Psi(\tau)$ with the distribution
\begin{equation} \label{eq:BM}
\Psi(\tau) = \frac{\tau_0^\mu}{\Gamma(\mu)}\frac{\mathrm{e}^{-\tau_0/\tau}}{\tau^{1+\mu}}
\end{equation}
where $\Gamma(\cdot)$ is the gamma function, $\tau_0$ sets the cutoff
of the distribution at short absorption times, and $\mu\ge 0$ is the
so-called persistence exponent \cite{Bray:2013ha}.  At short times,
this distribution has a exponential cutoff that is a consequence of
finite speed processes and, as our present results indicate, it is
asymptotically dominated by a power-law tail. Note that the
distribution in Eq.~\eref{eq:BM} is exact for a one-dimensional
Brownian motion in the semi-infinite line, with $\mu=1/2$ and
$\tau_0=x_0^2/2D$, where $x_0$ is the starting position and $D$ the
diffusion coefficient \cite{Redner:2007ua}. The light yellow dashed
curves in Fig.~\ref{fig:fptd} correspond to the fit to
Eq.~\eref{eq:BM}, and the fitting parameters are reported in
Table~\ref{tab:fitparam}.

The  sharper  than   exponential  cutoff  at  short   times,  for  the
transmission across a polygonal channel, could be 
{smoothed  out}  by  considering  a   higher  power  of  the  exponent
$\tau_0/\tau$. Moreover,  the short times oscillations  are the result
of the ballistic fronts and  ultimately, to the underlying skeleton of
the deterministic periodic orbits. Although Eq.~\eref{eq:BM} is not expected to be
an exact description  for our systems, the  comparison is physically
appealing, given its broad applicability. For instance,
taking   $\mu=1-1/\alpha$,   Eq.~\eref{eq:BM}  provides   a   good
approximation  for  the  distribution  of  first-arrivals  for  L\'evy
flights, with L\'evy index $\alpha$,
$1  <  \alpha <  2$  \cite{Chechkin:2003if}.   Also, for  $\mu=1-H$,
Eq.~\eref{eq:BM}  is  a  reasonable  approximation  for  the  FPTD  of
fractional  Brownian  motion  with  Hurst  index  $H$,  where  $0<H<1$
\cite{HANSEN:2011hj}   (for   a   more   detailed   discussion   about
Eq.~\eref{eq:BM} see \emph{e.g.} Ref.~\cite{MejiaMonasterio:2011co}).

Furthermore, the distribution in Eq.~\eref{eq:BM} is normalised, but
for $0\le\mu<1$ its first moment, the MFPT, does not exist. For
$1\le\mu<2$, the MFPT exists but the variance of $\tau$ diverges, an
so on.  Observing the decay of the tails of $\Psi(\tau)$ in
Table~(\ref{tab:fitparam}), we realise that for the rational polygonal
channels, the MFPT, {\em i.e.}\ the mean time a particle takes to
escape from the channel, diverges.  This has important consequences
for the transmission through a finite rational polygonal channel: even
for a finite number of randomly initialised particles for which the
MFPT to escape typically exists, a large variability in the absorption
times is to be expected, yielding a meaningless MFPT
\cite{Mattos:2012ib}.  This calls for a characterisation of the
transmissibility that is not based on the mean escape time.  In
contrast, for irrational polygonal channels, the tail of $\Psi(\tau)$
decays faster than $\tau^{-2}$, namely with $1<\mu<2$ so that the MFPT
exists but the variance of the escape time diverges.

To study the variability in the escape times, we consider the
similarity index $\omega$ studied in the context of first-passage
times in Ref.~\cite{MejiaMonasterio:2011co}. Consider a set of $N$
trajectories, and let $\{\tau_j\}$ denote the set of escape times
corresponding to these trajectories. Then the similarity index is
defined by the ratio between any of the escape times, say $\tau_j$
with $j\in\{1,\ldots,N\}$, and mean escape time averaged over the set
of $N$ trajectories:
\begin{equation} \label{eq:sim} \omega_N =
  \frac{1}{N}\frac{\tau_j}{\langle \tau \rangle_N} \ ,
\end{equation}
where $\langle \tau \rangle_N  = \frac{1}{N}\sum_{j=1}^N \tau_j$. This
quantity probes  the escape time  of any given trajectory  relative to
the  ensemble  average  over  $N$  trajectories.  The  domain  of  the
similarity index is  $\omega_N \in [0,1]$ and is a  random variable if
the  escape   times  $\{\tau_j\}$  are  random   variables.   A  low
variability means that $\tau_j \approx \langle \tau \rangle_N$ for all
the $N$ trajectories, yielding $\omega_N \approx \frac{1}{N}$.

Here  we explore  the  simplest expression  for  $\omega$, namely  for
$N=2$.  Consider   two  trajectories  with  randomly   chosen  initial
conditions.  The first trajectory exits the channel at a time $\tau_1$
while the  second trajectory  does it  at a  time $\tau_2$.
Then, we have $\omega_2=\tau_1/(\tau_1+\tau_2)$
\footnote{Alternatively one can define
  $\omega_2=\tau_2/(\tau_1+\tau_2)$, which ultimately indicates that
  the probability distribution $P(\omega_2)$ is symmetric over its domain.}.  When all
trajectories are absorbed at approximately the same time,
$\omega_2 \approx 1/2$.  When variations are large, {\em i.e.}\
$\tau_1 \gg \tau_2$ or $\tau_2 \gg \tau_1$ , $\omega_2$ attains a
value close to $0$ or to $1$. For clarity in what follows we refer to
$\omega_2$ simply as $\omega$.

\begin{table}[!t] 
\centering
\begin{tabular}{|c|c|c|c||l|c|}
    \hline 
  polygon & horizon & $\alpha$    & $d$ & $\tau_0$ & $\mu$ \\
    \hline 
    \hline 
rational & finite &   $\frac{\pi}{2}$ & $\delta y/2$ & $60$ & $0.536$ \\
rational & infinite &   $\frac{\pi}{2}$ & $2\delta y$ & $80$ & $0.498$ \\
irrational & finite &    $\left(\varphi-\frac{3}{5}\right)\frac{\pi}{2}$ & $\delta y/2$ & $400$ & $1.38$ \\
irrational & infinite &    $\left(\varphi-\frac{3}{5}\right)\frac{\pi}{2}$ & $2\delta y$ & $960$ & $1.2176$ \\
    \hline 
\end{tabular}
\caption{Parameters $\tau_0$ and $\mu$ obtained from the fit to
  Eq.~\eref{eq:BM}, shown in Fig.~\ref{fig:fptd} as dashed light yellow
  curves. Note the striking difference between rational and irrational
  channels.}
\label{tab:fitparam}
\end{table}

The distribution of $\omega$ for the FPTD of Eq.~\eref{eq:BM} is
given by \cite{MejiaMonasterio:2011co}
\begin{equation} \label{eq:Pw}
  P(\omega) = \frac{\Gamma(2\mu)}{\Gamma^2(\mu)}
  \omega^{\mu-1}(1-\omega)^{\mu-1} \ .
\end{equation}
In the insets  in Fig.~\ref{fig:fptd} we show the  corresponding numerically
computed  distributions $P(\omega)$  as  solid black  curves. For  the
rational  polygonal channel,  $P(\omega)$ has  a characteristic  ``U''
shape,   with  most   probable   values  of   $\omega$   at  $0$   and
$1$. Strikingly,  the mean $\langle\omega\rangle  = 1/2$ is  the least
probable  value of  the  distribution. Such  large  variations of  the
absorption times are a consequence of the broadness of the FPTD.

For the irrational polygonal channels, the distribution of $\omega_2$
{is maximal at} the mean $\langle\omega\rangle = 1/2$ for the
infinite horizon, and is shaped like a ``M'' for the finite horizon.
While the variability of the absorption times is markedly smaller than
for the rational polygons, variations remain wild, as indicated by the
almost flat profile of $P(\omega)$ near the center of the domain.

While the description in terms  of $P(\omega)$ is only qualitative, it
clearly  {reveals}  a  remarkable {difference}  between  the  rational
polygon  with $\alpha=\pi/2$  for  which the  MFPT  diverges, and  its
irrational counterpart for which the MFPT exist but its variability is
large to the extend that it does not characterise the escape time of a
typical  trajectory.  In the  next  section  we will  generalise  this
observation to any rational and the irrational polygons.

As a final observation, we note that with
the exception of the irrational polygon with finite horizon,
Eq.~\eref{eq:Pw}, shown as the dashed light yellow curves in the
insets of Fig.~\ref{fig:fptd}, is a reasonably good approximation of
the numerically obtained $P(\omega)$.  However, the ``M'' shape
obtained for the irrational polygon with finite horizon cannot be
described by Eq.~\eref{eq:Pw}, but is characteristic of a FPTD with
power-law tails with an additional exponential cutoff at large $\tau$
\cite{MejiaMonasterio:2011co}, and its origin deserves further
investigation.

\subsection{Survival probability}
\label{sec:surv}

In the previous Section we have seen that it is not possible to use
the mean first-passage time to characterise the transmission through
finite polygonal channels. In this Section we explore a global
characterisation of the transmission process in terms of the survival
probability. The survival probability $S(t)$ is the fraction of
particles that have not yet escaped {at} time $t$, and is related to
the FPTD through
\begin{equation} \label{def:surv}
\Psi(\tau) = -\frac{\mathrm{d}S(\tau)}{\mathrm{d}\tau} \ .
\end{equation}

In our settings, the trajectories initially prepared in the central
cell of the channel can indeed be physically seen as a short pulse of
particles released in the center of the channel.  By measuring the
time at which the particles escape {from} the channel, we compute the
survival probability $S(t)$ and characterise the transmission of the
initial pulse, by defining a channel escape time $\tau_\epsilon$ as
the hitting time at which the survival probability decays to a value
{smaller} than {a given} $\epsilon>0$, namely
\begin{equation} \label{eq:tauesc}
\tau_\epsilon = \inf\left\{t | S(t) < \epsilon\right\} \ .
\end{equation}
We hence define the channel escape time as the time at which a
fraction $(1-\epsilon)$ of the total particles has escaped from the
channel.  \jo{Being an} observable integrated over a finite time it
is not affected by the divergence of the moments of $\Psi(\tau)$
and, interestingly, is one of the observables frequently measured in
short pulse-response experiments \cite{Marin:2019ug}.

Indeed, the inverse problem of {inferring} details of the geometry
{from} the survival probability is a standard problem of scattering
processes.  In recent years {it has been investigated} in the
so-called stochastic billiards, that are random processes derived from
deterministic billiards \cite{Feres:2007it,Comets:2010eg}.

\begin{figure}[!t]
    \centering
    \includegraphics[width=0.8\textwidth]{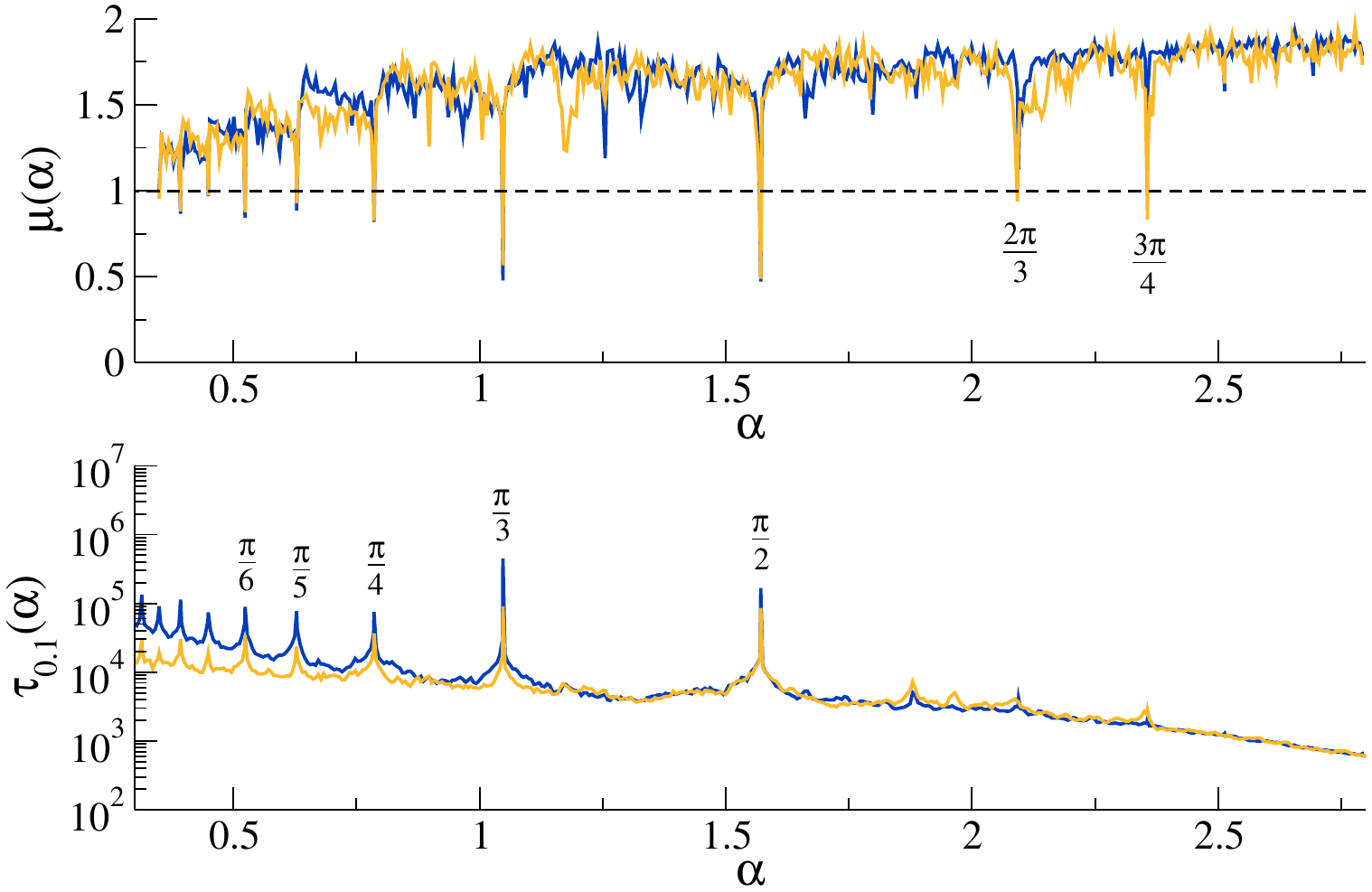}
    \caption{Top panel: \jo{Persistence exponent $\mu$ of the FPTD
        of Eq.~\eref{eq:BM}} as a function of the aperture angle $\alpha$, for a finite
      $d=\delta y/2$ (dark blue curve) and infinite $d=2\delta y$
      (light yellow curve) horizons. The horizontal dashed line
      indicates the value of $\mu$ above which the MFPT does not
      exist.  Bottom panel: Corresponding escape time $\tau_\epsilon$
      of Eq.~\eref{eq:tauesc} for $\epsilon=0.1$, as a function of the
      angle $\alpha$.  The results in this figure were derived from
      the survival probability obtained for each $\alpha$,
       in a channel made of $L = 101$ cells.}
    \label{fig:taub_nu_alpha}    
\end{figure}

From Eqs.~\eref{eq:BM} and \eref{def:surv}, the survival probability
decays asymptotically as \jo{$S(\tau) \sim \tau^{\mu-2}$}. This is in
contrast to polygonal channels with non parallel walls, for which
$S(\tau)$ was instead found to decay as a stretched exponential
\cite{Alonso:2004bf}.

To explore the transmission in polygonal channels, we computed the
statistics of the first-passage time as a function of the aperture
angle $\alpha$, for finite and infinite horizons. For all values of
$\alpha$, the survival probability (and thus the FPTD) decay
asymptotically as a power-law of $\tau$. From a fit to power-law, we
obtained the exponent $\mu$ for the decay of the FPTD in
Eq.~\eref{eq:BM}, as a function of $\alpha$, with the results shown in
the top panel of Fig.~\ref{fig:taub_nu_alpha}.

Strikingly,  for  all simple  rational  polygons  with aperture  angle
$\alpha = \pi/n$,  and $n$ \jo{an} integer, the decaying exponent  of the FPTD
is \jo{$\mu < 1$},  meaning that for all simple rational  polygons the mean
first-passage time diverges. The  same is observed for $\alpha=2\pi/3$
and  $3\pi/4$.  In contrast,  other  generic  values, including  the
irrational polygons,  have \jo{$1 < \mu < 2$}  meaning that the  MFPT exists,
but  the variance  of  the  escape time  diverges.  This result  holds
irrespectively of whether the channel's horizon is finite or infinite.
We argue that the divergence of the MFPT is characteristic of rational
polygons, and the reason why we  do not observe this for all \jo{rational 
aperture angles $\alpha_R$,} lies   in  the  infinite  numerical   accuracy  needed  to
distinguish rational and irrational angles.

In the lower panel of Fig.~\ref{fig:taub_nu_alpha}, we show the
channel escape time $\tau_\epsilon$ for $\epsilon=0.1$, namely the
time at which the $90$\% of the initial particles have left the
channel. The escape time is {about} an order of magnitude larger for
simple rational polygons than for irrational ones, indicating that in
rational polygonal channels transport drastically slows down,
{presumably because of} the dense skeleton of periodic orbits
characterising the channels' dynamics \cite{Boshernitzan:1998kd}.

\section{Conclusions}
\label{sec:concl}

We have addressed the nature of mass diffusion in a family of open
polygonal channels with parallel walls, analysing the statistics of
particle displacement. While parallel channels have been subject to
implicit or partial examination in previous works, the key questions
of those studies concern the implications of normal transport in the
absence of chaotic dynamics. This paper has instead a broader scope
and has \jo{enabled} insight into the fine tuning of transport
properties with the details of the channel's geometry.

The  probability distribution  function of  the particle  displacement
$P(\Delta  x,t)$ is  a  combination of  two  contributions: a  leading
scale-invariant  contribution with  exponential  tails describing  the
short  fluctuations of  the displacement,  and a  sub leading ballistic
contribution, describing  the evolution of the  ballistic fronts.  The
scale  of the  leading  contribution is  $\Delta  x/t^\gamma$, with  a
characteristic scale  $\gamma<1$ that depends  on the geometry  of the
channel.

The  ballistic  front exists  independently  of  the geometry  of  the
channel.  In  particular, it  is present  irrespective of  whether the
horizon is finite  or infinite.  The geometry determines  the speed of
the front.  For an aperture angle  $\alpha = \pi/2$ the speed seems to
saturate  when the  horizon is  finite, namely  for $d<\delta  y$, and
converges    to    its    upper    bound    $v_{\mathrm{bal}}=1$    as
$d\rightarrow\infty$. We  argue that  this behaviour  is qualitatively
shown  by  arbitrary  values  of the  aperture  angle.   Moreover,  we
speculate that  the ballistic  front is due  to the  trajectories that
initially are in a close  neighbourhood of the shortest periodic orbit
of the polygonal billiard. That  would explain not only the dependence
of the speed of the ballistic front on the geometry of the channel but
would  also shed  light  on  the strong  variability  observed on  the
transport properties. This  question will be \jo{addressed}  in a forthcoming
research.

In addition to the geometry  dependent ballistic fronts, a free-flight
ballistic front of unit speed appears,  as soon as the horizon becomes
infinite.   However, the  fraction of  such fastest  trajectories does
necessarily decrease in  time, since it approaches the  set of exactly
horizontal trajectories. With infinite  horizon it would be possible in
principle to describe the transport  in polygonal channels in terms of
the           single            big           jump           principle
\cite{Burioni:2010jp,Vezzani:2020ki}.  However,  our  results  suggest
that the sub leading ballistic  contribution vanishes rapidly enough to
affect the properties of transport.

For arbitrary geometry mass transport is super-diffusive, meaning that
the MSD scales faster than linear in time. Polygons with rational
aperture angle $\alpha$ show a MSD with a power scaling $\gamma$ that
increases smoothly with $\alpha$, in contrast to irrational angles for
which the power scaling is roughly constant. At large aperture angles
both families converge to a ballistic behaviour as
$\alpha\rightarrow\pi$. It still remains to be understood whether the
coefficient for the MSD scaling of irrational parallel billiards,
which appears to be almost constant throughout the range of the
aperture angles we simulate, is an indication of a form of structural
stability for these kinds of systems.

Exploring  the  spectrum of  higher  moments  of the  displacement  we
obtained  that   transport  is   characterised  by   strong  anomalous
diffusion, with  moments higher  than $p^\star$  scaling ballistically
and lower  moments exhibiting the  same scaling  as the MSD.   For the
geometries investigated,  the threshold  moment is $p^\star>2$  in all
cases, meaning  that the contribution  of the ballistic fronts  to the
MSD vanishes sufficiently  fast. In any case, the  spectrum of moments
of  the displacement  suggest that  the transport  in these  polygonal
channels  belong  to  the  universality class  recently  described  in
Ref.\cite{Vollmer:2021ut}.

Borrowing concepts from stochastic dynamical systems, we have analysed
the transmission of particles through finite channels under the light
of first-passage time statistics. The distribution of FPT to leave the
channel is dominated at short times by the ballistic fronts, and
asymptotically decays as power-law, with a scaling that is determined
by the channel's geometry. Fat tailed FPT distributions have important
consequences on the variability of escape times. Strikingly, our
numerical results suggest that when the aperture angle $\alpha$ is a
simple rational, all moments of the distribution of FPT diverge, in
particular the \jo{mean first passage time}. For other angles the mean escape time
exists, but its value does not represent the typical transmission due
to the divergence of the escape time variance. The striking difference
between statistics of the FPT for rational and irrational channels is
interesting, pointing to highly dissimilar dynamics associated to
heterogeneous structures of periodic and ballistic trajectories for
the two cases.  Notably, the consequences for experiments involving
pulses and assemblies of particles are non trivial.

Findings discussed in  the present work do not only  pertain to highly
idealised systems.  Our investigation  of transport properties, and in
particular their dependence  on the geometry of the  channel, may help
us elucidate the properties of realistic systems, for example those of
transport and solution kinetics, or a variety of settings that involve
diffusion and propagation  of pulses, cells and  tracers.

\ack CMM does also acknowledges financial support from the Spanish
Government grant PGC2018-099944-B-I00 (MCIU/AEI/FEDER, UE).  LR
acknowledges that the present research has been partially supported by
MIUR grant Dipartimenti di Eccellenza 2018-2022 (E11G18000350001). JO
and FF finally acknowledge financial support from the Australian
Research Grant DP180101512 and thank Swinburne University of
Technology for the generous allocation of CPU time on the OzSTAR HPC cluster.

\section{References}
\bibliographystyle{unsrt} 
\bibliography{polygon}

\end{document}